\documentclass[aps,prd,preprint]{revtex4-1}

\usepackage{amsmath}
\usepackage{blkarray}
\usepackage{multirow}
\usepackage{mathtools}
\usepackage{graphicx}

\usepackage{graphics}
\usepackage{graphicx}
\usepackage{epsfig}
\usepackage{color}
\usepackage{xspace}
\begin{document}

\newcommand{\bw}{\texttt{BayesWave}\xspace}
\newcommand{\BayesLine}{\texttt{BayesLine}}
\newcommand{\LALInference}{\texttt{LALInference}}
\newcommand{\cWB}{\texttt{cWB }}
\newcommand{\Msun}{${\mathrm{M}_\odot}$}
\newcommand{\Mc}{{\mathcal{M}}}
\newcommand{\data}{d}
\newcommand{\h}{{\bm{h}}}
\newcommand{\n}{{\bm{n}}}
\newcommand{\params}{{\boldsymbol \theta}}
\newcommand{\intp}{{\boldsymbol \lambda}}
\newcommand{\extp}{{\boldsymbol \Omega}}
\newcommand{\Hyp}{{\mathcal{H}}}
\newcommand{\Bay}{{\mathcal{B}}}
\newcommand{\Sig}{{\mathcal{S}}}
\newcommand{\Gli}{{\mathcal{G}}}
\newcommand{\Odd}{{\mathcal{O}}}
\newcommand{\IFO}{{\rm IFO}}
\newcommand{\SNR}{{\mathrm{SNR}}}
\newcommand{\BSG}{{\ln\mathcal{B_{S,G}}}}
\newcommand{\BSN}{{\ln\mathcal{B_{S,N}}}}
\newcommand{\fd}{\dot{f}_0}

\DeclareGraphicsExtensions{.pdf,.gif,.jpg}

\title{Bayesian reconstruction of gravitational wave bursts using chirplets}
\author{Margaret Millhouse}
\affiliation{eXtreme Gravity Institute, Department of Physics, Montana State University, Bozeman, MT 59717, USA}
\author{Neil J. Cornish}
\affiliation{eXtreme Gravity Institute, Department of Physics, Montana State University, Bozeman, MT 59717, USA}
\author{Tyson Littenberg}
\affiliation{NASA Marshall Space Flight Center, Huntsville AL 35812, USA}

\begin{abstract}
The LIGO-Virgo collaboration uses a variety of techniques to detect and characterize gravitational waves. One approach is to use templates - models for the signals derived from Einstein's equations. Another approach is to extract the signals directly from the coherent response of the detectors in LIGO-Virgo network. Both approaches played an important role in the first gravitational wave detections. Here we extend the \bw analysis algorithm, which reconstructs gravitational wave signals using a collection of continuous wavelets, to use a generalized wavelet family, known as chirplets, that have time-evolving frequency content. Since generic gravitational wave signals have frequency content that evolves in time, a collection of chirplets provides a more compact representation of the signal, resulting in more accurate waveform reconstructions, especially for low signal-to-noise events, and events that occupy a large time-frequency volume.
\end{abstract}

\maketitle

\section{Introduction}

The first gravitational wave signal recorded by the advanced LIGO detectors~\cite{TheLIGOScientific:2014jea}, GW150914, was detected by multiple search pipelines~\cite{GW150914}, some that directly reconstruct the signal using a coherent wavelet representation~\cite{Abbott:2016ezn}, and others that use waveform models, or templates, derived from general relativity~\cite{TheLIGOScientific:2016pea}. Both types of analysis play an important role in gravitational wave (GW) detection and characterization: the modeled searches are the most sensitive to the types of signals they are targeting, while the direct reconstruction methods are sensitive to a wider variety of GW sources. Both types of analysis also play an important role in understanding the physical properties of the sources. The template based analysis provides a mapping between the shape of the waveform and the physical properties of the source, allowing us to infer that GW150914 was produced by the collision of two black holes, each $\sim 30$ times the mass of the Sun~\cite{TheLIGOScientific:2016wfe}. The greater flexibility of the wavelet-based reconstruction approach was used to test for deviations from the predictions of general relativity~\cite{TheLIGOScientific:2016src}.

While all of the GWs detected so far have been from compact binary systems, there exist other GW sources that are not as well modeled: for example core-collapse supernovae; post-merger oscillations of hypermassive neutron stars; magnetar flares; and pulsar glitches.  There is also the possibility that LIGO/Virgo could detect gravitational waves from a completely new and unpredicted sources.  In addition to having good models for well understood sources, it is also crucial that we are ready to both detect and characterize any possible astrophysical signal.

One analysis technique that has been widely used in LIGO to detect and reconstruct GWs with minimal assumptions is the \bw algorithm \cite{BW}.  Several studies \cite{BWII, BWIII, Bence} have shown \bw's capability to robustly distinguish between real astrophysical signals and transient noise artifacts (glitches) that are known to occur in the detectors, and to faithfully reconstruct waveforms from simulated signals.  It does this by reconstructing the detector data using a sum of Morlet-Gabor sine-Gaussian wavelets. The number of wavelets used is determined by the data, with more complicated signals ({\it i.e.} those having more structure in time-frequency space) needing more wavelets.  Because Morlet-Gabor wavelets have variable shapes in time-frequency space they are generally able to fit waveforms well, but there are several avenues for improving the fidelity of the waveform reconstructions. One is to modify the prior on the wavelet placement - such as using a ``clustering prior''~\cite{BW} which assigns greater probability to regions in time-frequency that are close to other wavelets. Another is to change the wavelet model. Here we investigate the use of  ``chirplets'' - modified sine-Gaussian wavelets with linear frequency evolution~\cite{Chirplets}. The motivation for using chirplets is that the frequency content of GW signals typically evolves with time~\cite{ChassandeMottin:2005jp,Candes:2006yv,Pai:2007ms,ChassandeMottin:2010js,Mohapatra:2011mx}. The chirplets can model both increasing and decreasing frequency evolution.

\section{Chirplet frame}\label{section:chirplets}

The Morlet-Gabor sine-Gaussian wavelets currently used by \bw form an over-complete basis, technically a {\em frame}~\cite{Kovacevic2008AnIT}, that can reconstruct any possible signal.  They have a simple analytic representation in the Fourier domain, making it easy to search over the time of arrival, and allowing for efficient calculation of the likelihood function.  But there may be other frames that are able to reconstruct signals more efficiently.

In choosing a new wavelet frame for \bw, we consider what types of gravitational-wave signals we might detect.  Many astrophysical sources of GWs have frequency content that evolves in time, most notably mergers of compact binary objects.  Because of this, we might expect that using a frame function that itself includes frequency evolution could better reconstruct GW signals.  The simplest way to incorporate changing frequency into \bw is to add a linear frequency evolution to the Morlet-Gabor wavelets, producing a function known as a chirplet~\cite{Chirplets}.

\begin{figure}[ht]
	\includegraphics[width=0.45\textwidth]{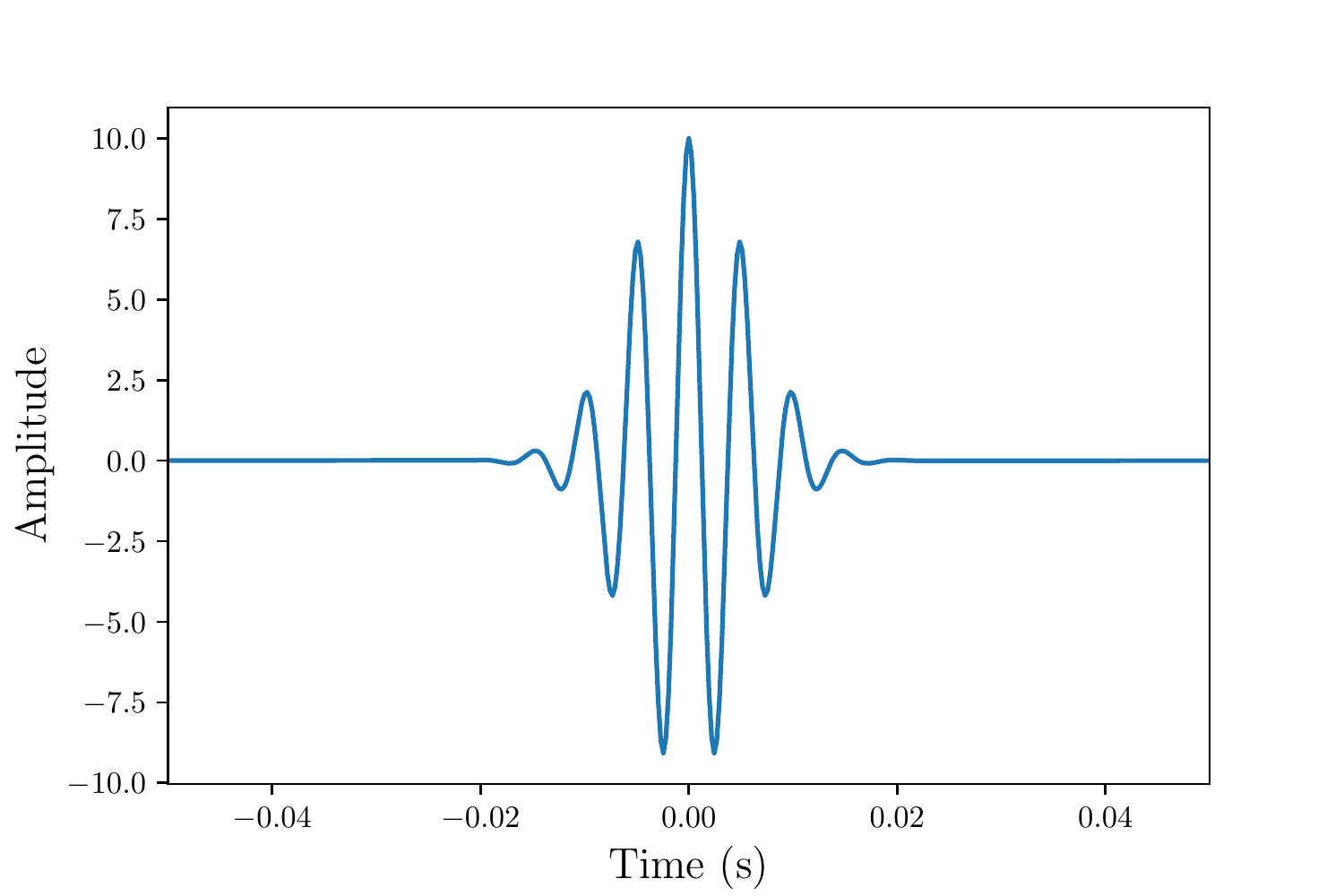}
	\includegraphics[width=0.45\textwidth]{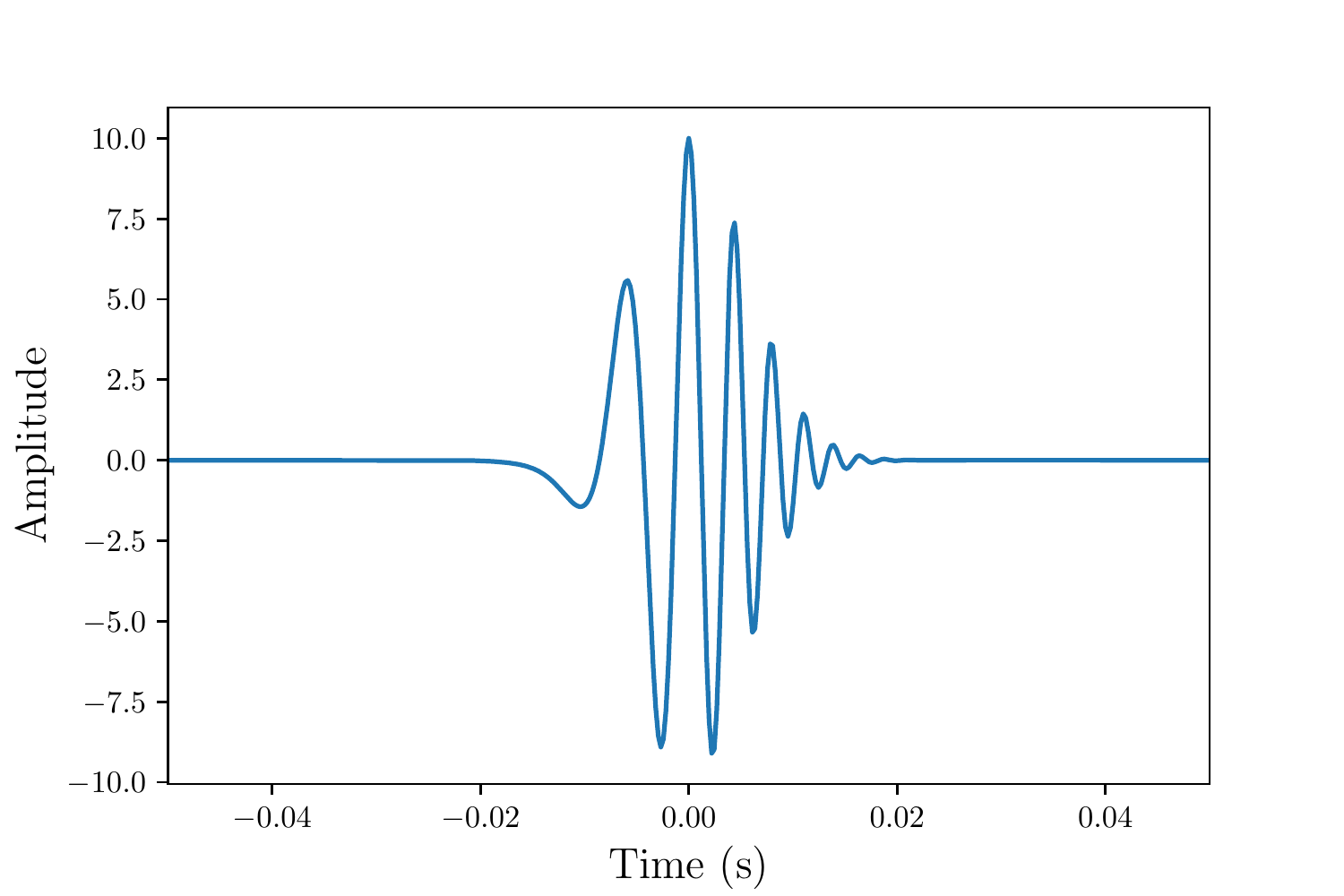}
\caption{Examples of a wavelet (left) and chirplet (right) in the time domain.  For both examples $f_0=200$, $t_0 = 0$, and $Q=10$.  In the chirplet example $\beta=0.8$}
\label{fig:TimeDomainEx}
\end{figure}

In the time domain, chirplets can be expressed as
\begin{equation}
\Psi(t;A,f_0,\dot{f}_0,Q,t_0,\phi_0)=Ae^{-\Delta t^2/\tau^2}\cos(2\pi f_0\Delta t + \pi \fd\Delta t^2+\phi_0)
\end{equation}
where $\tau = Q/(2\pi f_0)$ and $\Delta t = t-t_0$.  Here $\fd$ represents the frequency evolution and is the linear frequency derivative at time $t=t_0$. In the limit that $\fd=0$, this expression reduces to the expression for Morlet-Gabor wavelets.  Chirplets can have either $\fd>0$ (chirping), or $\fd<0$ (anti-chirping). Time domain plots of a chirplet with $\fd>0$ and a Morlet-Gabor wavelet ($\fd=0$) are shown in Figure \ref{fig:TimeDomainEx}.

The same characteristics that make Morlet-Gabor wavelets a good frame are also true for chirplets: they are continuous, occupy a small time-frequency volume, and can be expressed analytically in the Fourier domain, though with a slightly more complicated expression:
\begin{equation}
\begin{split}
\Psi(f;A,f_0,\fd,Q,t_0,\phi_0) = &\frac{A\sqrt{\pi} \tau}{2(1+\pi^2\beta^2)^{1/4}}e^{-\frac{\pi^2 \tau^2\Delta f^2}{1+\pi^2 \beta^2}}e^{-2\pi i f t_0}(e^{i(\phi_0+\delta-\pi^3\beta\tau^2\Delta f^2)/(1+\pi^2\beta^2)} \\
& +e^{-Q^2f/f_0}e^{-i(\phi_0+\delta-\pi^3\beta^2\Delta f^2)/(1+\pi^2\beta^2)})
\end{split}
\end{equation}
where $\Delta f = f-f_0$, $\delta = \frac{1}{2}\arctan(\pi\fd\tau^2)$, and we have introduced the dimensionless parameter $\beta = \fd\tau^2$.  For the remainder of this paper we will us $\beta$ as our chirp parameter.

\begin{figure}[ht]
	\centerline{\includegraphics[width=0.7\textwidth]{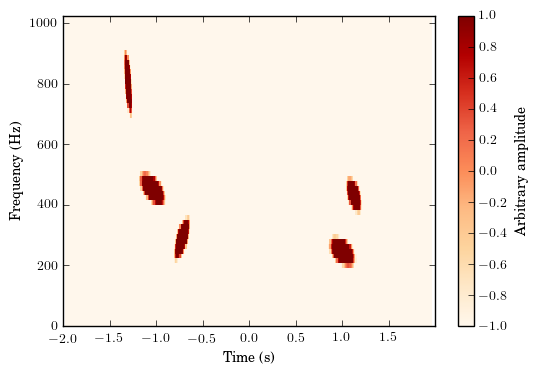}}
\caption{Spectrogram of chirplets with a range of central times, frequencies, $Q$ and $\beta$.}
\label{fig:SpecEx}
\end{figure}

In time-frequency space, wavelets can be represented by ellipses whose principle axes are aligned with the time and frequency axes.  Similarly, chirplets can be represented by a tilted ellipse. The equation for the ellipse is
\begin{equation}
(1+\pi^2\beta^2) x^2 + \pi^2 y^2 - 2\pi^2 \beta xy = 1
\end{equation}
where we have introduced the dimensionless variables $x=\Delta t/\tau$ and $y= \tau \Delta f$. In terms of these new coordinates, the ellipse is tilted with respect to the time axis by the angle
\begin{equation}\label{Eq:EllipseLengths}
\theta = \frac{1}{2}\, {\rm arctan} \left(\frac{2 \pi^2 \beta}{\pi^2(1-\beta^2)-1}\right)
\end{equation}
The ellipse has area $1+{\cal O}(\beta^4)$. A spectrogram of a variety of chirplets is shown in Figure \ref{fig:SpecEx}. 

\section{Methods}\label{section:methods}

Bayesian inference requires the specification of a likelihood and prior, and a method to compute the posterior distribution and model evidence. Since the replacement of sine-Gaussian wavelets by chirplets only adds one new parameter to the frame functions, the implementation is almost identical to the original \bw algorithm~\cite{BW}. In our study we started with the version of \bw used in the second advanced LIGO observation run, which differs from the original in the choice of priors, and in some of the proposal distributions used to evolve the MCMC algorithm.

\subsection{Priors}
We use uniform priors on $Q$ ($Q\in[0.01,40]$) and $\phi_0$ ($\phi_0\in[0,2\pi]$).  We also use a uniform prior on $f_0$ and $t_0$ over the time-frequency volume being analyzed.

The prior on the amplitudes of the individual wavelets (chirplets) is given as a prior on the SNR of the individual frame functions.  For an individual wavelet or chirplet, the SNR is estimated as:
\begin{equation}
\SNR\equiv 4 \int \frac{|\Psi(f;A,f_0,\fd,Q,t_0,\phi_0)|^2}{S_n(f)} df \simeq\frac{A\sqrt{Q}}{\sqrt{2\sqrt{2\pi}f_0S_n(f_0)}},
\end{equation}
where $S_n(f)$ is the one-sided power spectral density of the noise. The prior on the SNR for wavelets in signal model is the same as described in Refs.\cite{BW,BWII}:
\begin{equation}
p(\SNR) = \frac{3\SNR}{4\SNR_*^2\left(1+\frac{\SNR}{4\SNR_*}\right)^5}
\end{equation}
For the glitch model, the prior on the SNR is 
\begin{equation}
p(\SNR) = \frac{\SNR}{2\SNR_*^2\left(1+\frac{\SNR}{2\SNR_*}\right)^3}
\end{equation}
where $\SNR_*$ is the SNR at which the distributions peak, empirically chosen to be $\SNR_*=5$.

For chirplets, we limited the tilt of the ellipse in $\tau$-scaled time-frequency space to be below 45 degrees, which acording to Eq. \ref{Eq:EllipseLengths} corresponds to $\beta = \pm \sqrt{1-1/\pi^2}\approx \pm0.95$. Physically this limit corresponds to roughly a doubling of the frequency across the duration of the chirplet. For larger values of $|\beta|$ the chirplets no longer provide a very compact time-frequency representation. We adopt a uniform prior on $\beta$ in the range $\beta\in[-\sqrt{1-1/\pi^2},\sqrt{1-1/\pi^2}]$. The prior on the number of frame functions is uniform in $N_w\in[0,20]$. 

\subsection{Simulated Data}
To test the performance of the chirplet frame, we will look at how faithfully simulated GW signals can be reconstructed, and the number of frame functions used in the reconstruction.  Our test data set consists of the of binary black hole merger signals, and unpolarized white noise bursts in simulated Gaussian noise at the aLIGO design sensitivity~\cite{Aasi:2013wya}.

The binary black hole data set is a system of two 50\Msun back holes with the waveform generated using the Effective One Body approximation \cite{EOB} over a range of SNRs. We choose binary black holes as a test waveform in part because these are examples of waveforms we know have frequency evolution and thus are somewhere we believe a frame with frequency evolution could be beneficial.  We also now know that GWs from black hole systems are detectable by LIGO, and we can likely expect more of these signals in the future.  In addition to the standard BBH waveforms, we also tested BBH waveforms that have been time reversed, so that the frequency \emph{decreases} over time. This set is used to demonstrate that the chirplet frame is good for general signals with time-frequency evolution, and is not specifically targeting BBH signals.

The second class of waveforms used are unpolarized white noise bursts (WNBs).  These waveforms serve as a good test for the chirplet frame because they contain complicated frequency structure, which does not evolve smoothly like the BBH signals.  These signals can sometimes present challenges for \bw because they are unpolarized, whereas the current implementation of \bw assumes that the signals are elliptically polarized. We will see that this mis-modeling throws-off our estimates of the fidelity of the reconstruction as a function of signal-to-noise ratio. 

In previous studies of unmodeled searches \cite{Abbott:2016ezn}, sine-Gaussian waveforms (SGs) have also been used as test cases.  However, we have already seen that though \bw can reconstruct SGs well, we are relatively insensitive to them in a search.  This is a natural result of the fact that the signal-to-glitch Bayes factor scales with the number of wavelets used.  For a sine-Gaussian signal we expect and indeed see that \bw typically uses only one wavelet to reconstruct SGs, so the Bayes factor scales only with SNR, making it more difficult to distinguish between signals and glitches.  As wavelets are chirplets in the limit that $\fd\rightarrow0$, we see the same behavior from chirplets and so do not consider SGs here.


\section{Results}\label{section:results}

The two metrics we will look at are the number of frame functions used ($N$), and the match between waveforms, which as is defined as 
\begin{equation}
  \label{eq:match}
M = \frac{(h|\bar{h})}{\sqrt{(h|h)(\bar{h}|\bar{h})}}
\end{equation}
where $\bar{h}$ is the injected signal, and $h$  is the recovered signal, and $(a|b)$ denotes the standard noise-weighted inner product~\cite{maggiore2008gravitational} evaluated across the detector network. Here we are using a simulated network consisting of the LIGO Hanford and Livingston observatories operating at design sensitivity.

\begin{figure}[ht]
\centering
\includegraphics[width=0.45\textwidth]{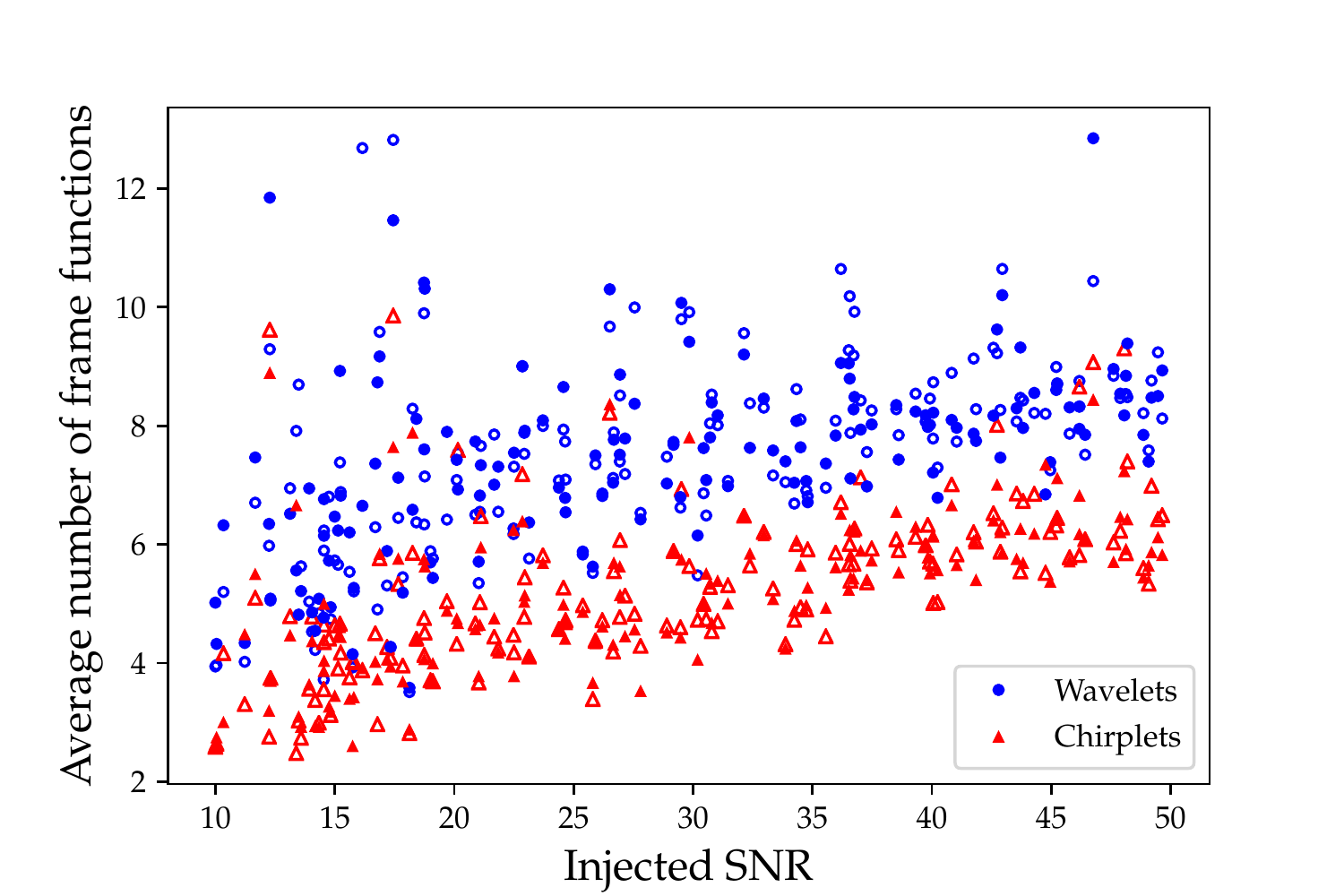}
\includegraphics[width=0.45\textwidth]{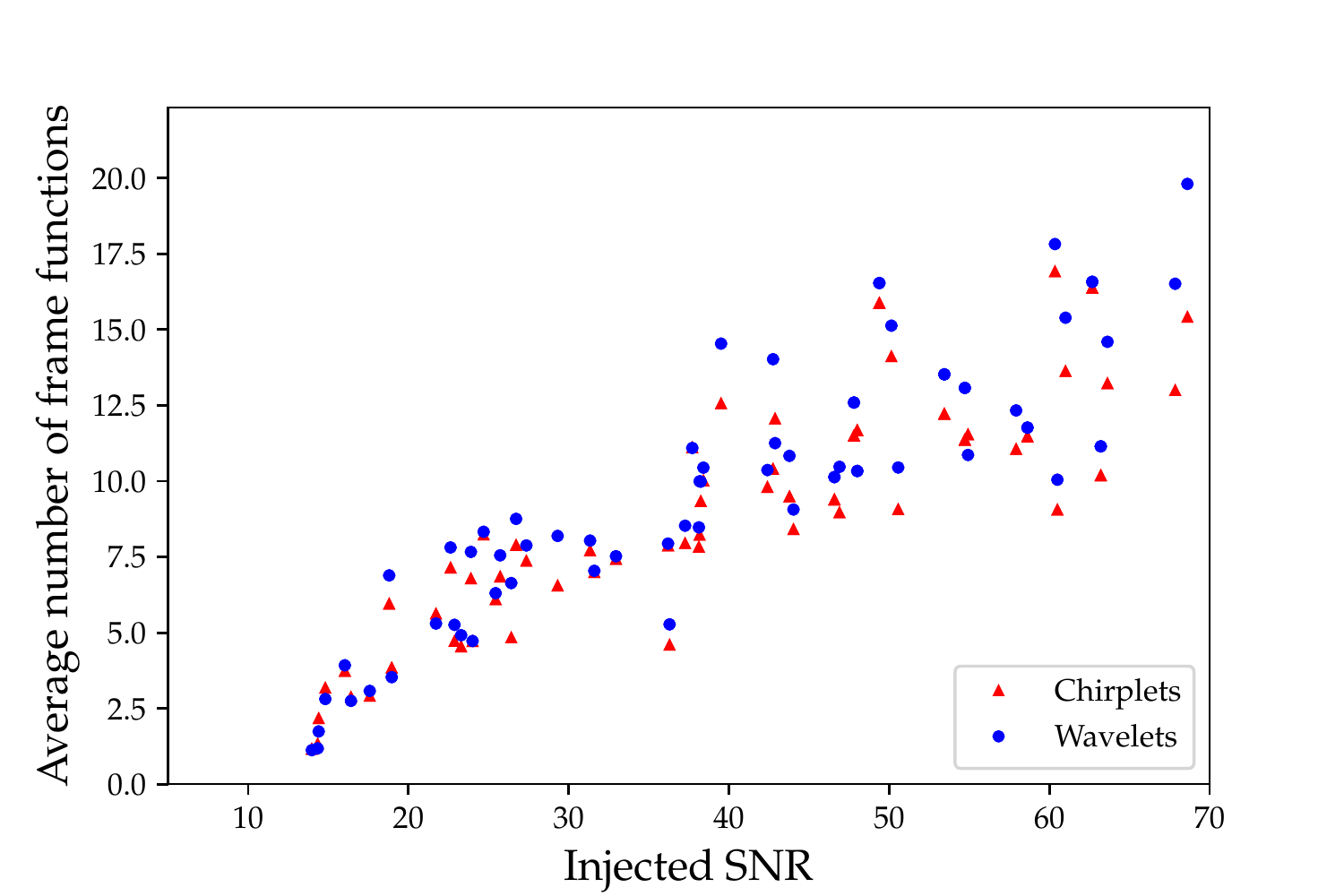}\\
\includegraphics[width=0.45\textwidth]{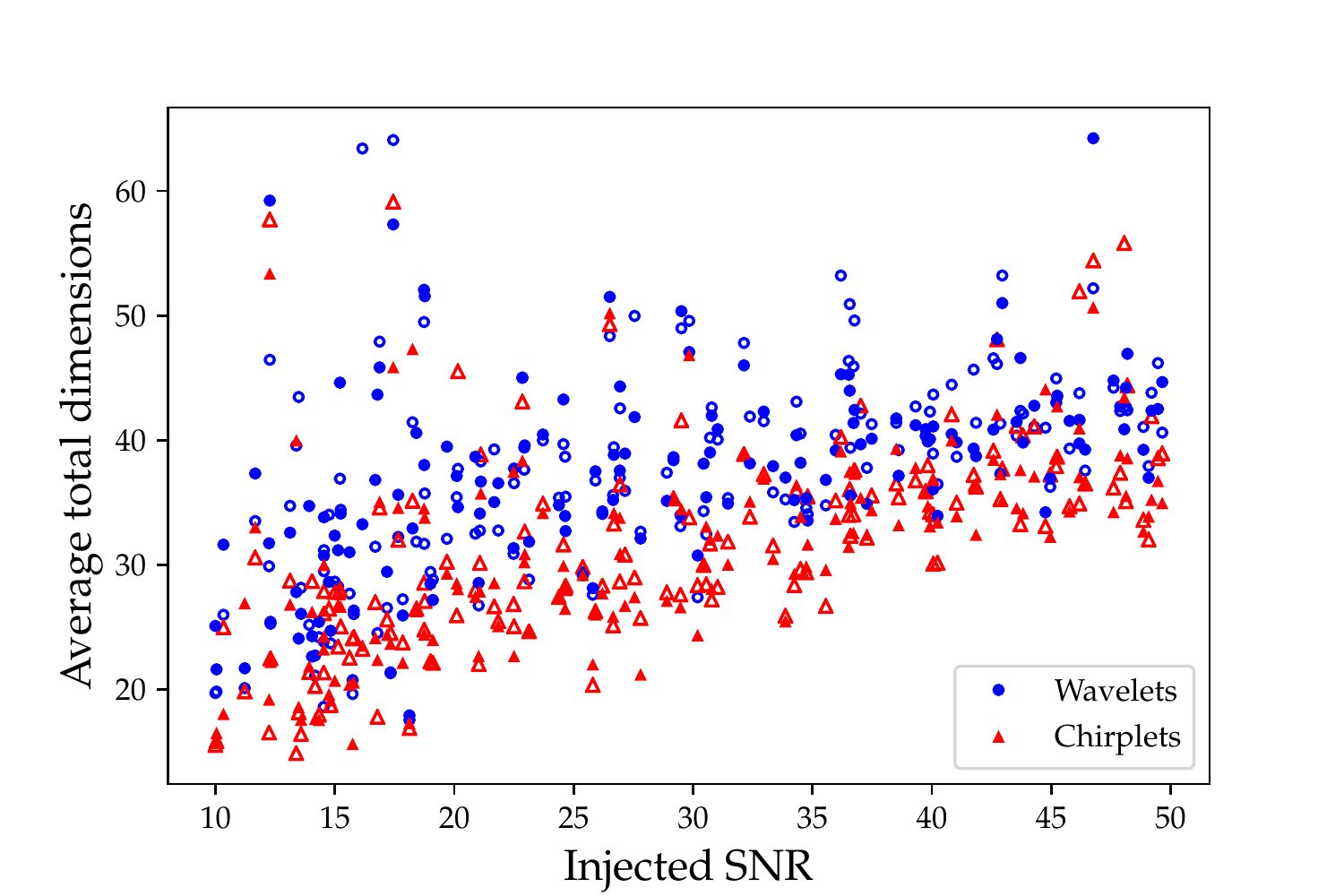}
\includegraphics[width=0.45\textwidth]{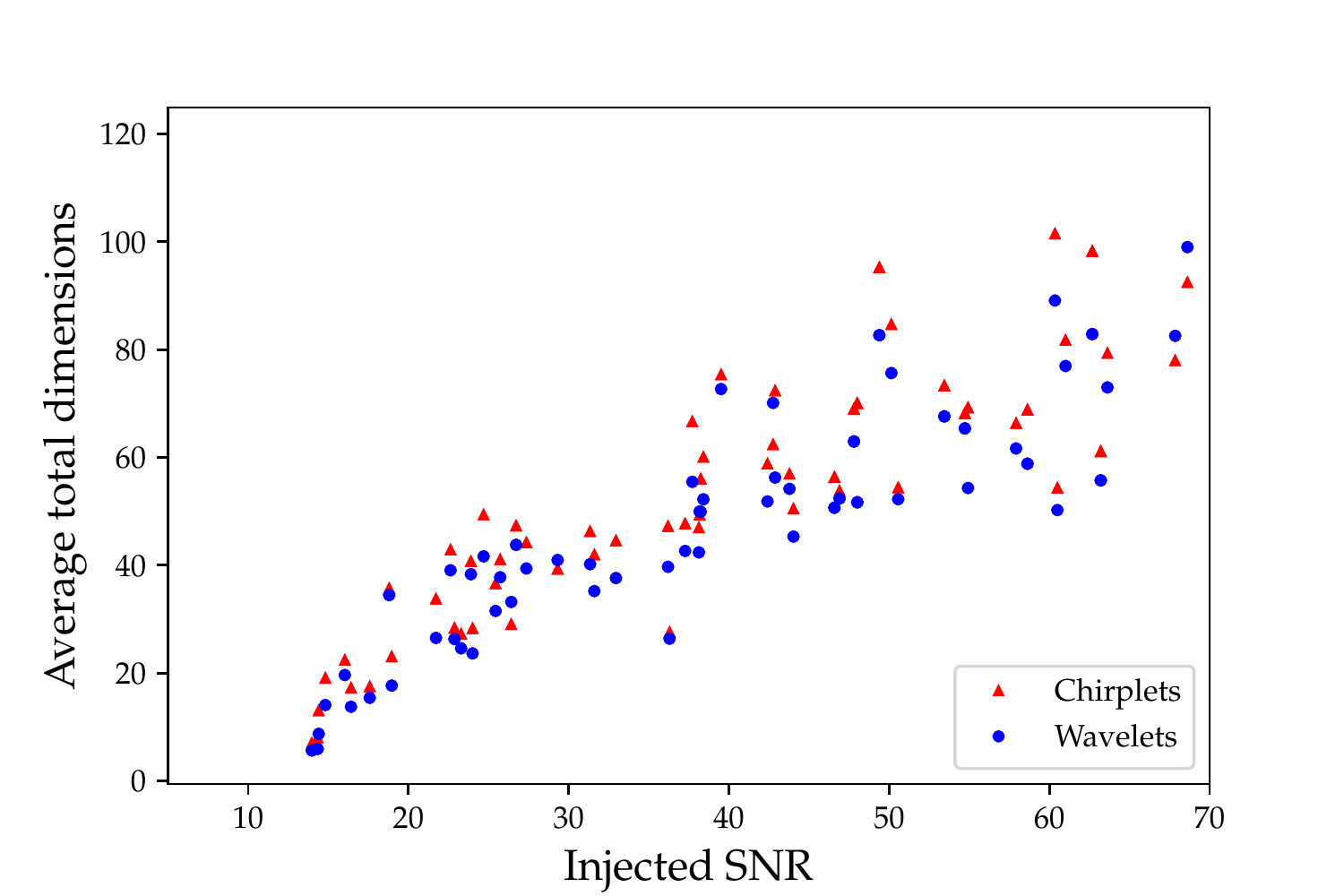}
\caption{The upper two panels show how the average number of frame functions (wavelets or chirplets) scales with the SNR. The lower two panels show the scaling for the total model dimension. The panels on the left are for 50\Msun-50\Msun BBHs, and the panels on the right are for white noise bursts. The filled markers in the panels on the left represent regular BBH injections, and hollow marker represent time-reversed BBH injections.}
\label{fig:snrvN}
\end{figure}

\subsection{Dimensionality}
\begin{figure}[ht]
\centering
\includegraphics[width=0.45\textwidth]{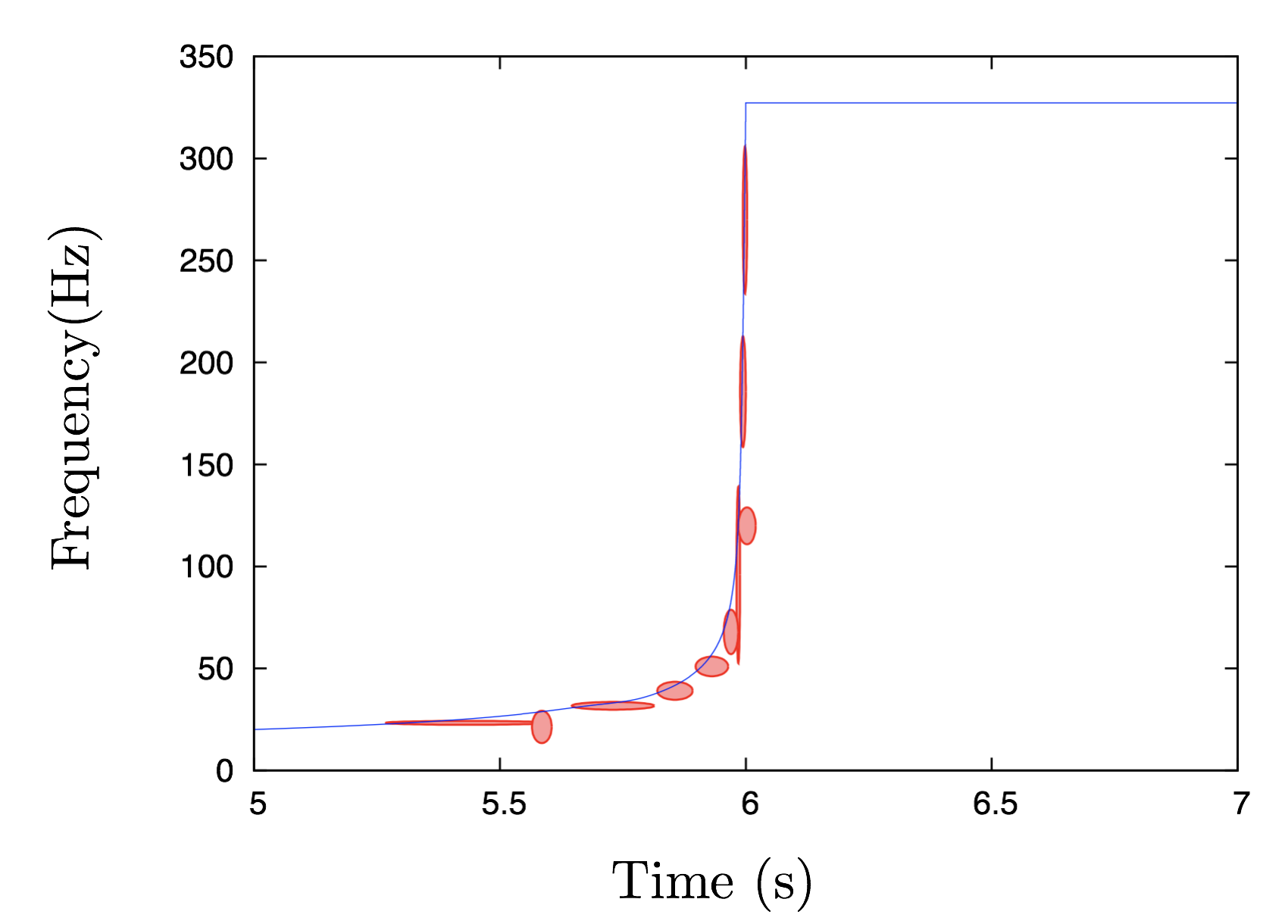}
\includegraphics[width=0.45\textwidth]{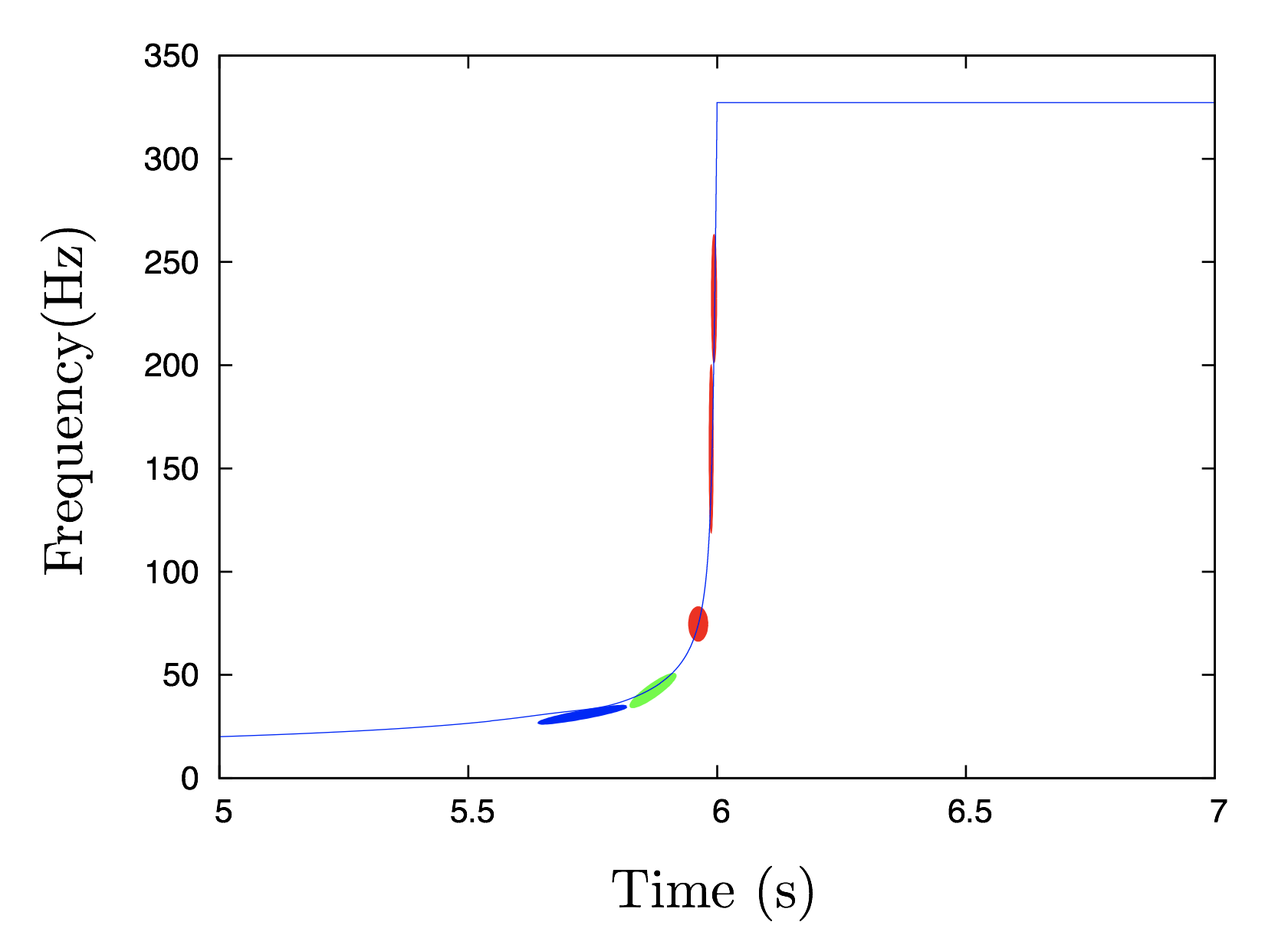}
\caption{An example of the wavelet frame (left) and chirplet frame (right) in action.  In this case we used a simulated BBH signal with component masses of 29\Msun and 30\Msun at an SNR of 35.  The solid line is the predicted $f(t)$ track and the colored ellipses are the wavelets or chirplets from a fair draw from the posterior distribution.}
\label{fig:track}
\end{figure}

The results for the average (mean) number of frame functions used for the BBH injection set and the WNB injection set are shown in Figure \ref{fig:snrvN}.  The difference is more apparent for the BBH injections, but we see that in general fewer chirplets are used than wavelets. This is as predicted-- the extra parameter, $\fd$ in the chirplet frame allows for fewer frame functions to be used in the reconstruction.  This implies that the extra flexibility of chirplets may make them preferable for waveform reconstruction, particularly at low SNRs.  A heuristic example of this can be seen in Fig.~\ref{fig:track}, where we see how the chirplets frequency evolution allows them to more closely follow the frequency evolution of the BBH signal.

We also see that, as shown in Ref.~\cite{BWII,BWIII}, the number of wavelets used is roughly linearly dependent on the SNR of the injected signal. In Ref.~\cite{BWII} this dependence was written as $N\approx1+\gamma\SNR$, but here we generalize this expression to:
\begin{equation}
N\approx \alpha + \gamma\SNR,
\label{eq:WaveletScaling}
\end{equation}
with the constants $\alpha$ and $\gamma$ being determined by the waveform morphology.  In practice this expression is only valid for sufficient large SNRs, otherwise the number of frame functions used drops rapidly to one (the minimum allowed number of frame functions in \bw's signal model). 

Using the results from the BBH injections, we preform a simple linear fit to find $\alpha$ and $\gamma$ for the wavelet and chirplet runs.  In both cases the slopes are very similar: $\gamma_{chirp} = 0.065$, $\gamma_{wave}=0.066$.  The starting number of frame functions though varies significantly: $\alpha_{wave} = 5.6$, and $\alpha_{chirp}=3.3$.  So while the number of frame functions used increases at a similar rate for both chirplets and wavelets, chirplets use reliably fewer frame functions.

For the WNB injections, we see that while for higher SNR injections slightly more wavelets tend to be used than chirplets, the difference is not nearly as striking as for BBH injections.  Again with a simple linear fit we see $\gamma_{chirp} = 0.21$, $\gamma_{wave}=0.23$, $\alpha_{chirp}=0.76$, and $\alpha_{wave}=0.53$, giving very similar slopes and starting points for both frame functions.

\begin{figure}[ht]
\centering
\includegraphics[width=0.45\textwidth]{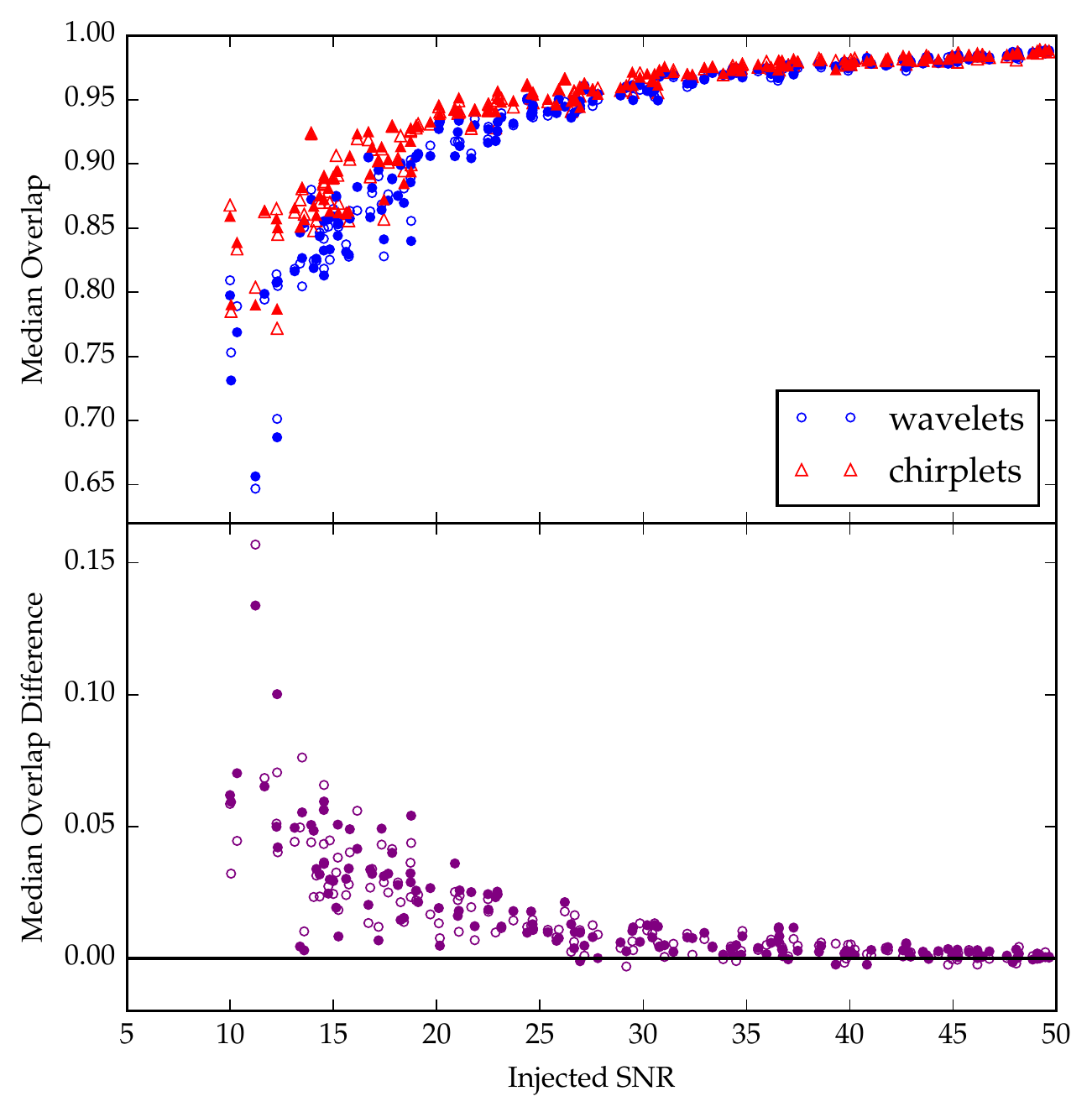}
\includegraphics[width=0.45\textwidth]{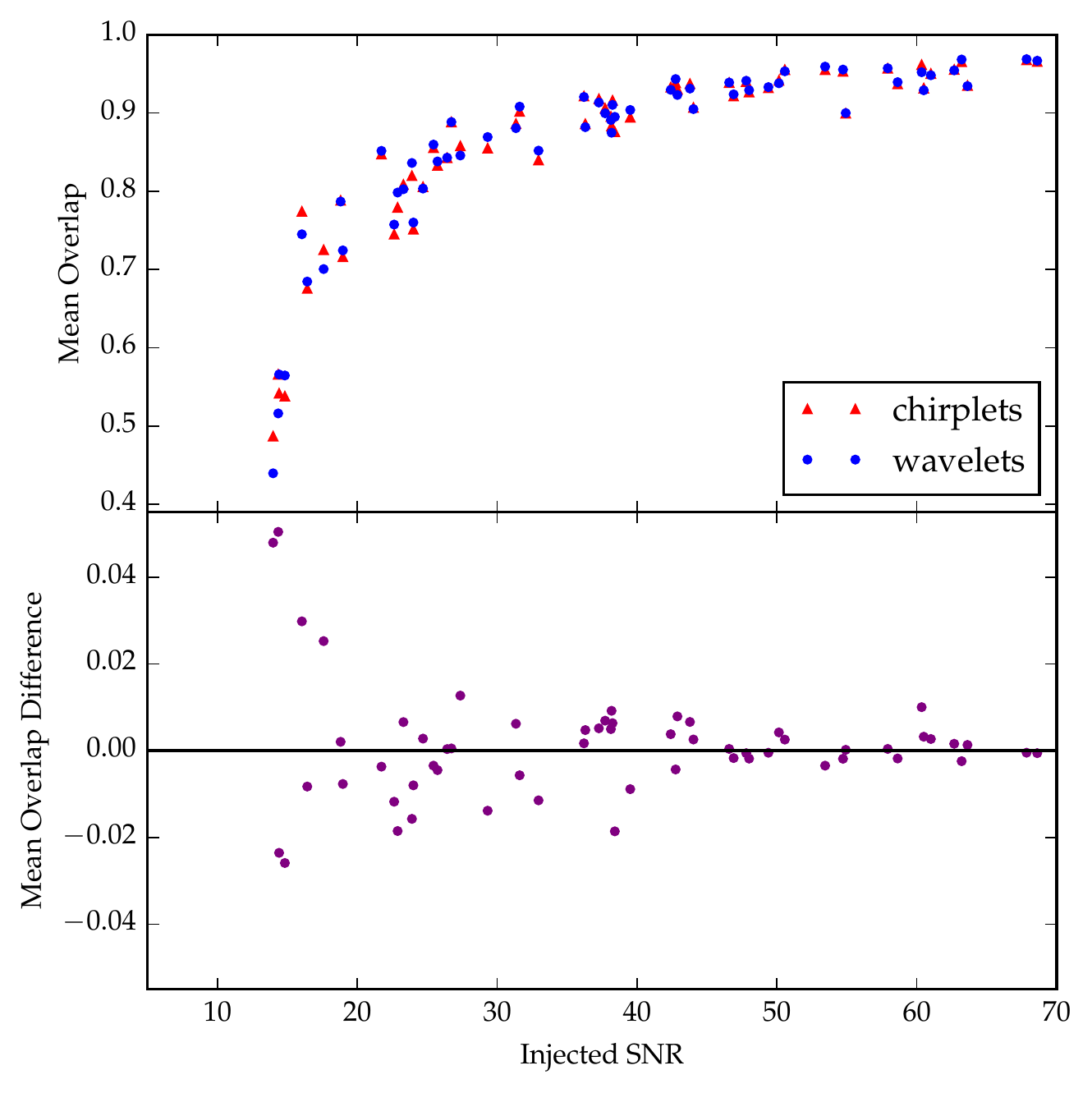}
\caption{The upper panels show the median match between the injected signal and reconstructed waveform versus SNR, while the lower panels shows the difference in the matches $\Delta M = M_\mathrm{chirplet}-M_\mathrm{wavelet}$. The plots on the left are for 50\Msun-50\Msun BBH signals, while the plots on the right are for unpolarized white noise bursts.}
\label{fig:snrvmatch}
\end{figure}

\subsection{Match}
As predicted, the chirplet frame generally uses fewer frame functions. To test how well the injected signal is recovered, we look at the match. Fig. \ref{fig:snrvmatch} shows the mean match between the injected and recovered waveforms for a set of two 50\Msun BHs in simulated aLIGO noise (left) and a set of WNBs (right) for a range of SNRs using either chirplets or wavelets as the frame function.  

\begin{figure}[ht]
\centering
\includegraphics[width=0.49\textwidth]{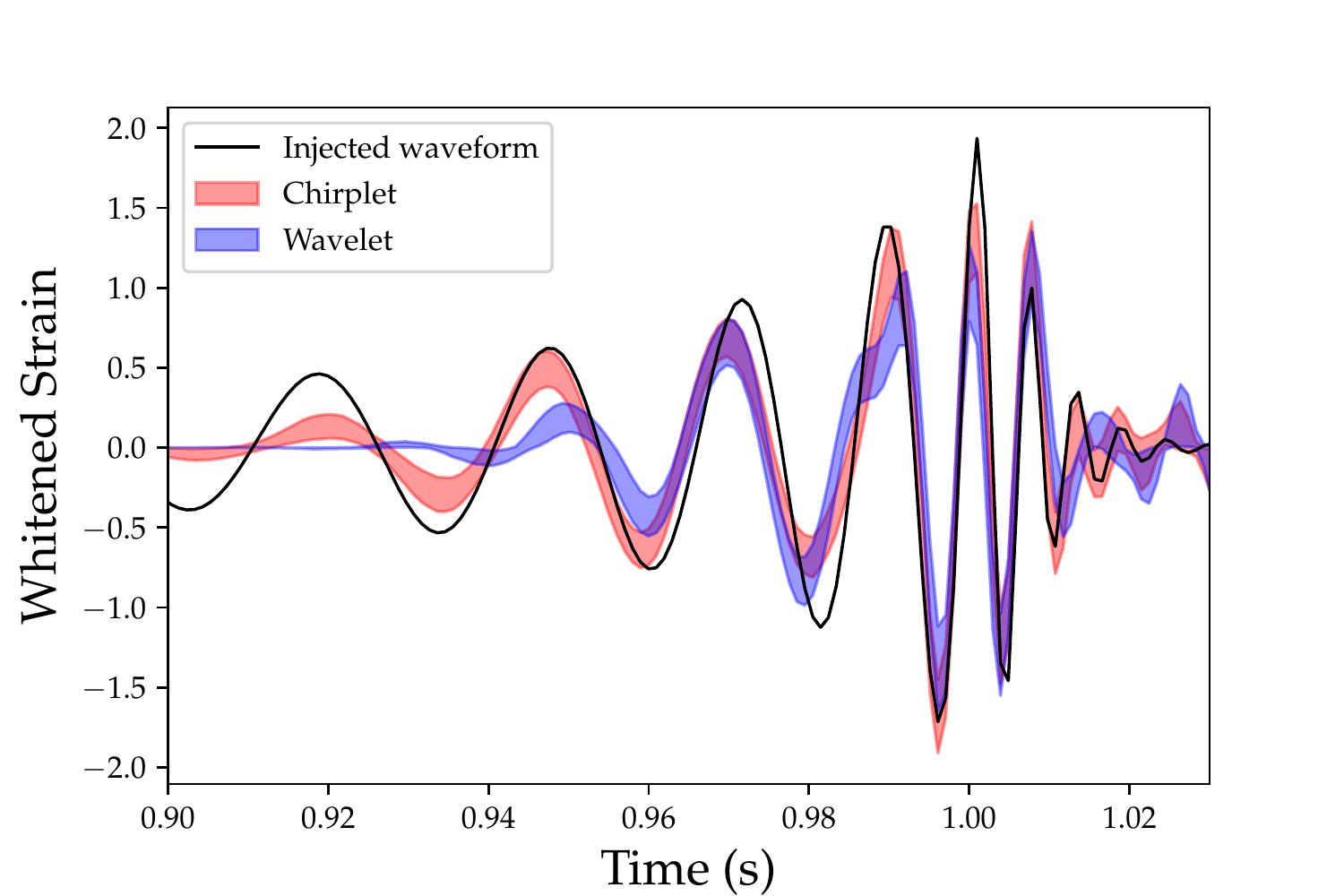}
\includegraphics[width=0.49\textwidth]{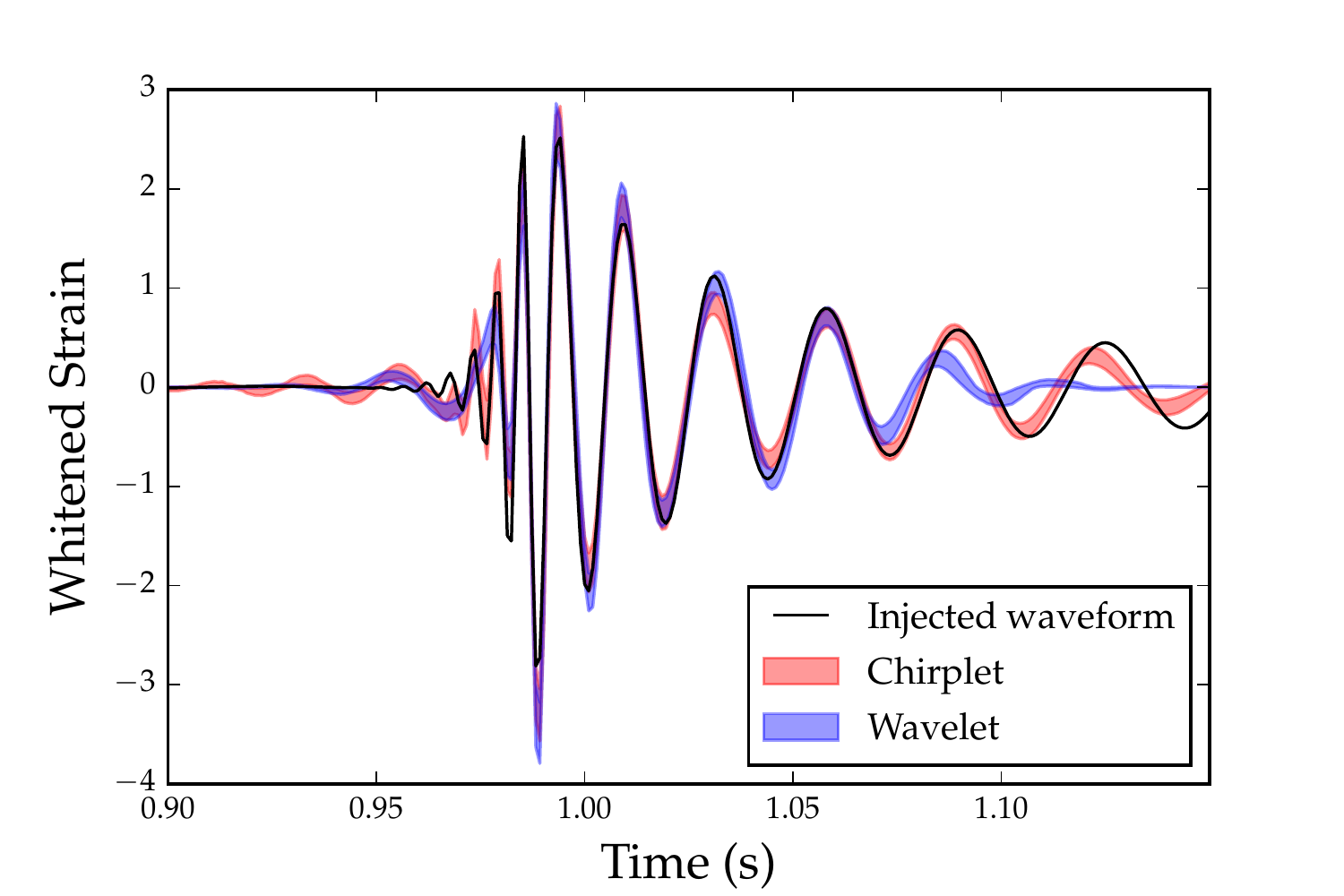}
\caption{Example of waveform reconstruction for a BBH, and a  time reversed BBH system (both at 50-50\Msun ). The red (blue) band shows the 50\% credible interval of the reconstructed waveform using the chirplet (wavelet) frame.  Both bases closely match the injected waveform (black) well in the higher power region, but chirplets are able to more accurately reconstruct the waveform in the regions with less power.  The BBH event was injected with SNR 11.2, and has a median match of 0.79 for the chirplet frame, and 0.66 for the wavelet frame. The time-reversed BBH event was injected with a network SNR of 10.25. The median network match for the chirplet frame is 0.91, and for the wavelet frame it is 0.87. }
\label{fig:antichirp}
\end{figure}

In the BBH case, for SNRs above about 25, the matches of the two different methods are comparable.  However at lower SNRs, we see that chirplets outperform wavelets, giving consistently higher matches.  This is important because low SNR events are more common that high SNR events, so small improvements in performance for low SNR signals can result in a large number of additional detections.  A particular example of a chirplet and wavelet reconstruction of a time-reversed BBH signal in shown Fig.~\ref{fig:antichirp}. The plots show the whitened strain, found by inverse Fourier transforming the Fourier domain signal $\tilde{h}(f)/\sqrt{S_n(f)}$. We see here that the chirplet frame manages to fit earlier and later parts of the signal.

\begin{figure}[h]
\centerline{\includegraphics[width=0.85\textwidth]{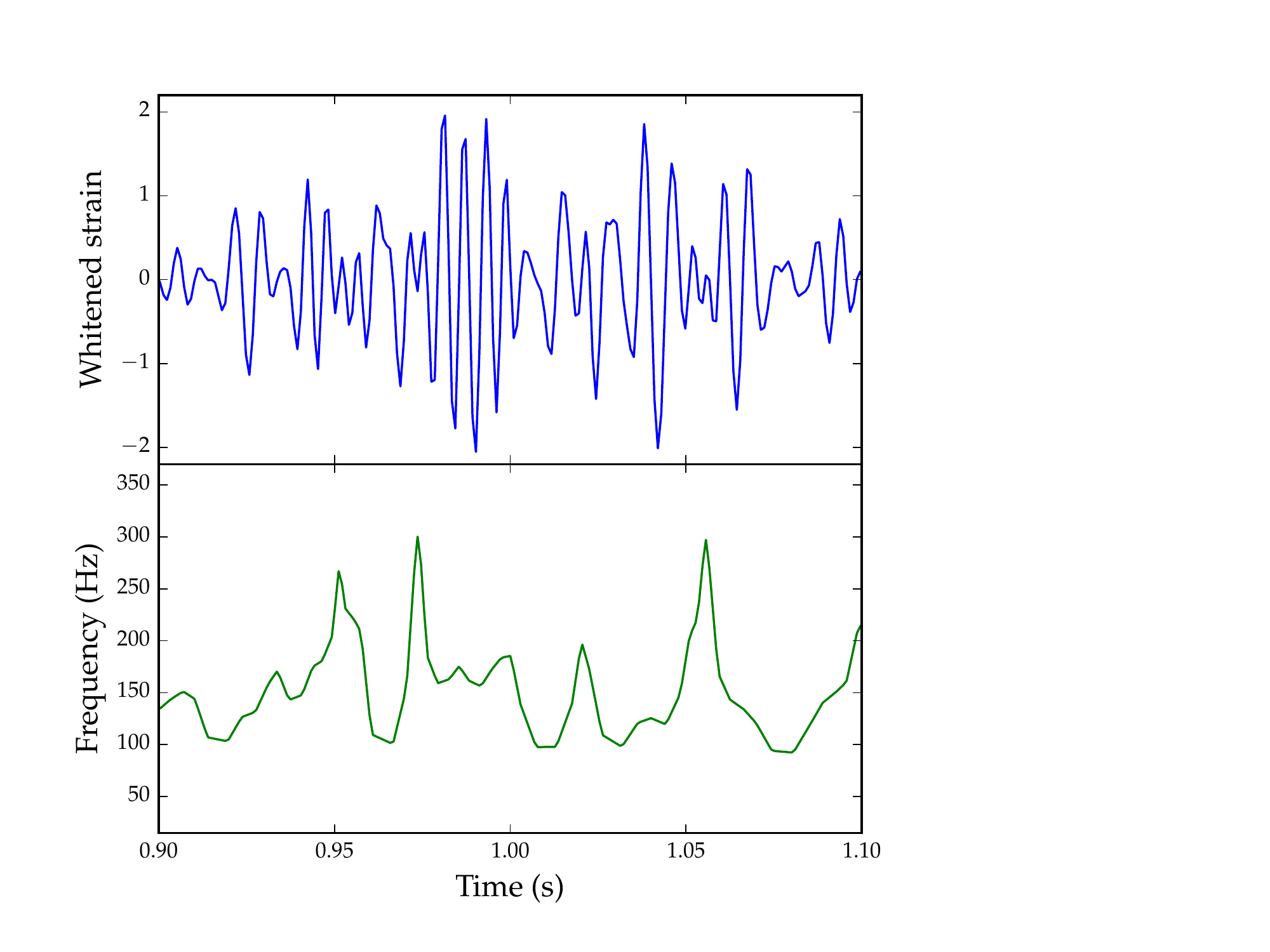}}
\caption{An example WNB waveform.  The whitened strain (top) and frequency as a function of time (bottom) show that WNBs are complicated signals with no well defined frequency evolution.}
\label{fig:WNBex}
\end{figure}

For the WNB injections, we see that the two frame functions perform about equally as well.  Previous injection studies with \bw have shown that WNBs can be difficult to to reconstruct.  One reason is that the WNBs are unpolarized, while \bw assumes an elliptical polarization.  WNBs also just have a very complicated, non-deterministic frequency evolution.  An example of the frequency evolution of a WNB is shown in Figure \ref{fig:WNBex}.  We see that for WNBs the frequency as a function of time changes rapidly from increasing to decreasing and back again.  Because the chirplets we use have only linearly increasing or decreasing frequency, the chirplet frame struggles to recover the fine details of signal. Thus we expect chirplets will provide the most benefit for signals with fairly smooth time-frequency evolution.

\begin{figure}[ht]
\centering
\includegraphics[width=0.45\textwidth]{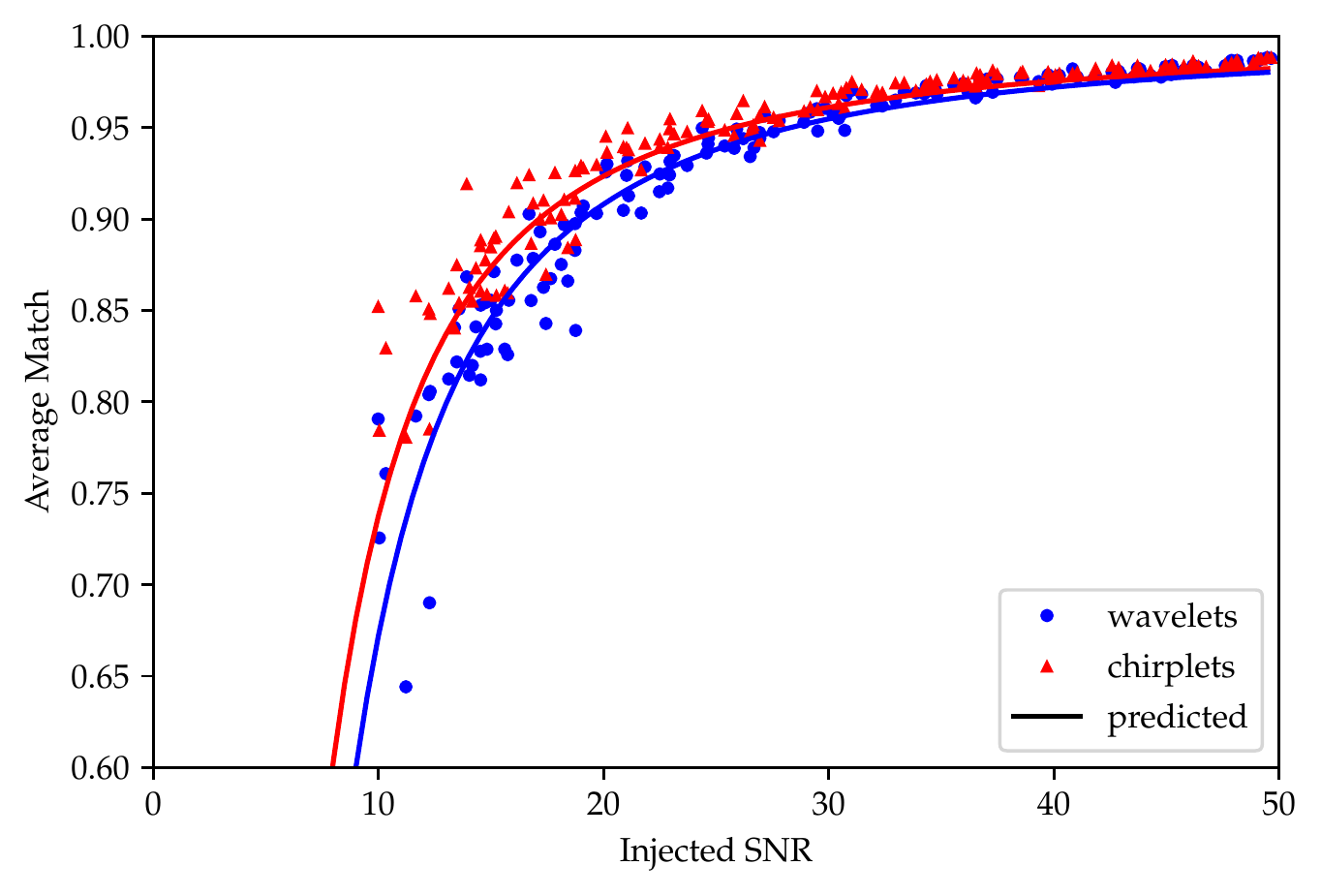}
\includegraphics[width=0.45\textwidth]{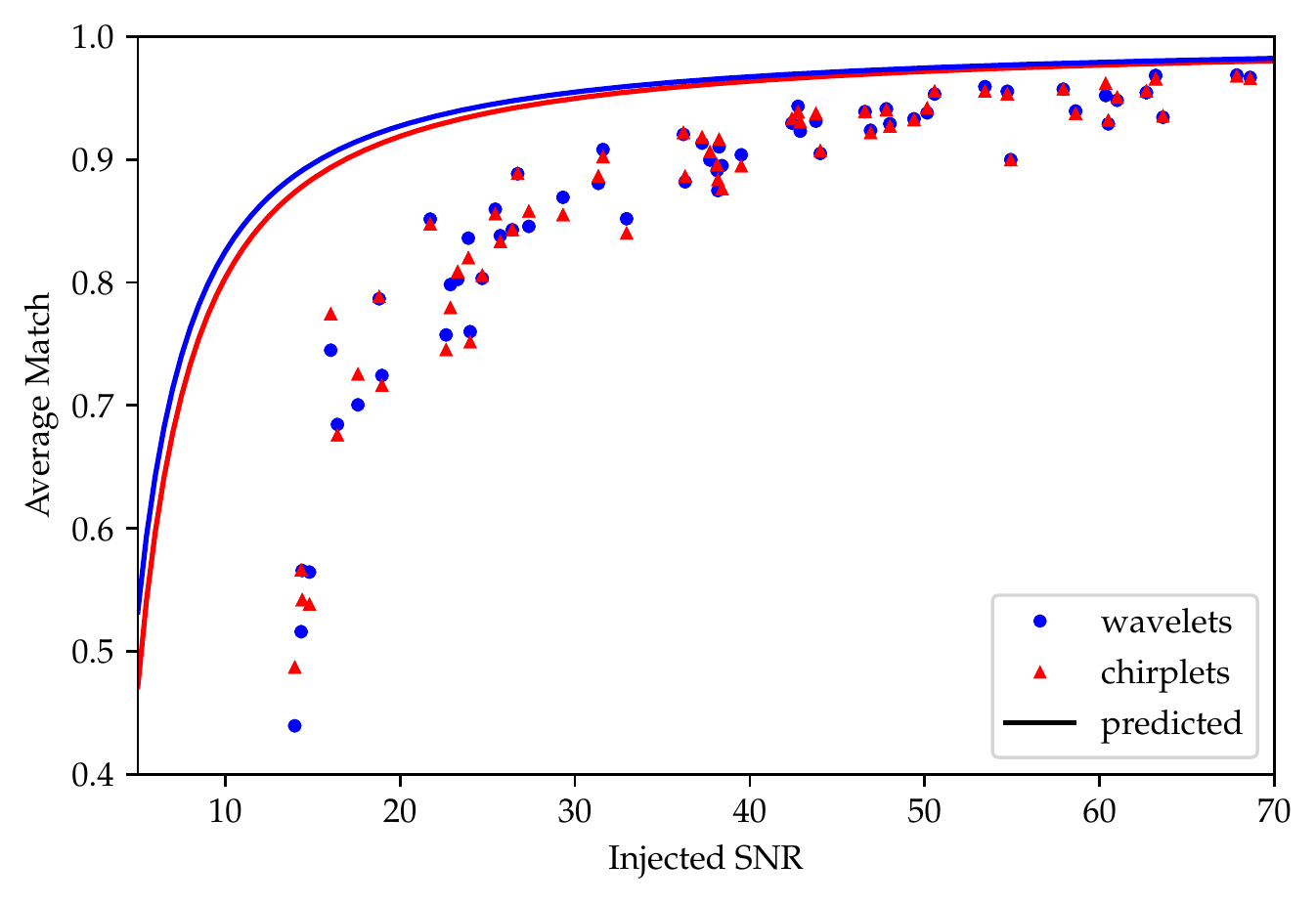}
\caption{The predicted match \ref{eq:EstimatedMatch2} plotted with the actual match for the set of BBH injections (left) and WNB injections (right).  The BBH injections generally follow the predicted match vs. SNR scaling, but the prediction overestimates the match for unpolarized WNB injections since the \bw signal model assumes elliptical polarization.}
\label{fig:SNRvMatch_predicted}
\end{figure}

We can also study what we theoretically would expect that matches to be for these injections.  For the match given in Eq.~\ref{eq:match}, we assume that the injected waveform $\bar h$ is dependent on parameters $\bar \lambda^i$, and the recovered waveform has parameters $\lambda^i$.  In the high SNR limit, the recovered and true injected parameters should be relatively consistent, or $\Delta \lambda^i = \bar \lambda^i - \lambda^i$ is small. Note that in this context the parameters are those of the wavelet/chirplet representation, and not, for example, the masses and spins of of the black holes. We can approximate the recovered waveform as:
\begin{equation}
h=\bar{h}+h_{,i}\Delta\lambda^i
\end{equation}
and $\Delta\lambda^i$ approximately follows the normal distribution:
\begin{equation}
p(\Delta\lambda^i)=\sqrt{\rm{det}(\Gamma/2\pi)}e^{-\Gamma_{ij}\Delta\lambda^i\Delta\lambda^j}
\end{equation}
where $\Gamma_{ij}=(h,_i|h,_j)$ is the Fisher information matrix.  We can expand our expression for the match, Eq.~\ref{eq:match} then to be \cite{OwenMatchFilter}
\begin{equation}
  \label{eq:expanded_match}
  M = 1-\frac{1}{2}\Delta \lambda^i \Delta \lambda^j\left(\frac{(h,_i|h,_j)}{(h|h)}-\frac{(h|h,_i)(h|h,_j)}{(h|h)^2}\right).
\end{equation}
Recognizing that the expected value of $\Delta \lambda^i \Delta \lambda^j$ is $E[\Delta \lambda^i \Delta \lambda^j]\approx \Gamma_{ij}^{-1}$ \cite{VallisneriFisher}, we find the expected match is:
\begin{equation}
E[M]\approx 1-\frac{D-1}{2\SNR^2}
\label{eq:EstimatedMatch}. 
\end{equation}
where $D$ is the dimension of the model. The minus one comes from the second term in Eq. \label{eq:expanded_match} removing the dependence on the amplitude of the signal. Note that this derivation assumes a templated search, and so should be thought of as more of a ``rule of thumb'' for this analysis.
Using the scaling for the number of wavelets in Eq. \ref{eq:WaveletScaling}, we have $D=N_p(\alpha+\gamma\SNR)+4$,
where $N_p$ is 5 for wavelets, and 6 for chirplets, and an additional 4 common extrinsic parameters (sky location, ellipticity and polarization angle).
The full expression for the predicted match is then:
\begin{equation}
E[M]\approx 1-\frac{N_p(\alpha+\gamma\SNR)+3}{2\SNR^2}
\label{eq:EstimatedMatch2}. 
\end{equation}
Figure \ref{fig:SNRvMatch_predicted} shows again the average match for the injected binary black hole signals, with the match predicted by Eq. \ref{eq:EstimatedMatch2}.  The recovered matches for the BBH injections follow the predicted match relatively well, however the recovered matches for the WNB injections are lower than the analytical prediction due to the signal model (polarized) not matching the simulated signals (un-polarized).

\subsection{Bayes Factors}
The \bw algorithm considers three distinct models: GW Signals + Gaussian noise $({\cal S})$; Noise transients (Glitches) + Gaussian noise $({\cal G})$; Gaussian noise $({\cal N})$; and computes evidence ratios, or Bayes factors, between the models. Here we investigate how the choice of frame impacts the Bayes factors between the models.

\begin{figure}[h]
\centering
\includegraphics[width=0.45\textwidth]{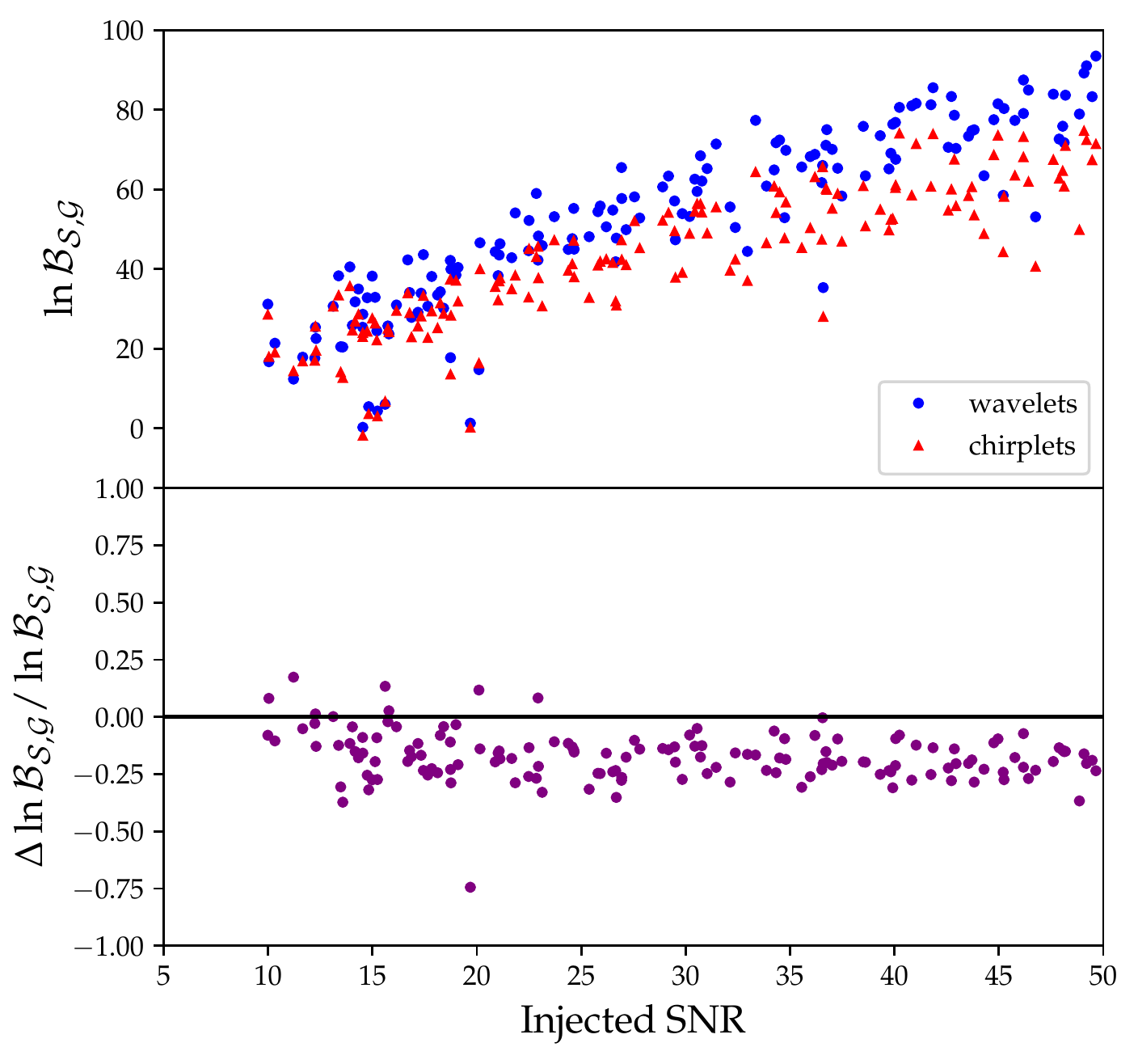}
\includegraphics[width=0.45\textwidth]{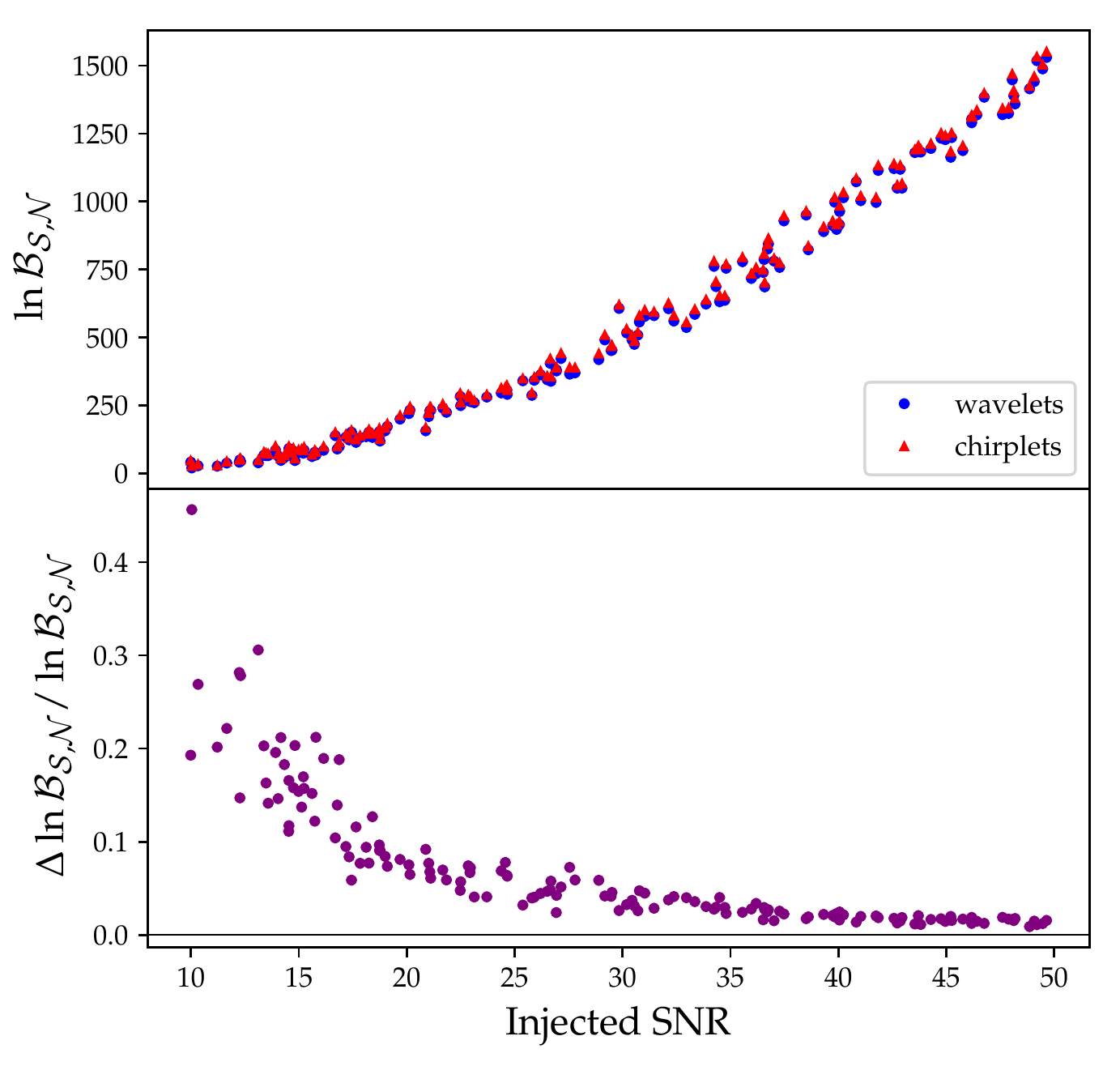}
\caption{The log signal-to-glitch (left) and signal-to-noise (right) Bayes factors for a set of $50-50M_\odot$ binary black holes using the chirplet and wavelet bases. The lower panel shows the difference in the log Bayes factors between the chirplet and wavelet frames, scaled by the wavelet frame log Bayes factor.}
\label{fig:BayesFactors}
\end{figure}

Figure \ref{fig:BayesFactors} shows $\BSN$ (left) and $\BSG$ (right) recovered using chirplets and wavelets for simulated binary black hole signals. We see that both bases return very similar $\BSN$, with the chirplet frame giving just slightly higher Bayes factors.  This is unsurprising since $\BSN$ scales with the recovered SNR, and chirplets are able to recover more SNR due to their ability to recover signals with higher fidelity.

The signal-to-glitch Bayes factors $\BSG$ show the opposite behavior, with the wavelet frame providing better separation between signals and glitches than the chirplet frame. This seemingly paradoxical result is due to the chirplet frame providing higher fidelity reconstructions using less parameters for {\em both} signals {\em and} glitches. Moreover, since the glitch model sees the signal in the individual detectors, which has lower signal-to-noise than the network response seen by the signal model, and since the chirplets outperform wavelets mostly at low SNR, the chirplet model boosts the evidence for the glitch model more than it boosts the evidence for the signal model, resulting in lower $\BSG$ than for the wavelet model. From the perspective of a search, where the goal is to separate signals from instrument noise, the wavelet frame outperforms the chirplet frame despite not doing as well at reconstructing signals. The same behavior was also seen when using the ``clustering prior''~\cite{BW}, which leads to higher matches, especially at low SNR, but worse separation between signals and glitches.

The reduction in the signal-to-glitch Bayes factor has prevented the clustering prior and the chirplet frame from being used in the current LIGO/Virgo analyses, despite the improvements they offer for signal and glitch reconstruction. We were forced to make this unsatisfactory choice because of the limitations in the specification of models. Going forward we plan to implement new models that do a much better job of separating signals and glitches, and that will not penalize models that do a better job of fitting low signal-to-noise features. One option is to modify the glitch model to be anti-coincident between detectors. This can be done by introducing a prior that disfavors placing wavelets at frequencies and times that are occupied by wavelets in the glitch models for the other detectors. 



\subsection{Discussion}

We have found that added flexibility offered by chirplet frame functions can reduce the overall model dimension, despite adding an additional parameter to each frame function, and improve waveform reconstruction, particularly at low SNRs. Limitations in the model selection approach that is currently used by \bw to distinguish between signals and glitches has so-far prevented the adoption of chirplets, but these limitations will soon be resolved. Ideally \bw should utilize a wide range of frame elements, including different types of wavelets and chirplets, and perhaps reduced-basis elements for black hole signals~\cite{Field:2011mf}. The optimal mix could then be dynamically selected via the trans-dimensional MCMC algorithm, hewing closer to our mantra ``model everything and let the data sort it out''.

\section{Acknowledgements}

MM and NJC appreciate the support of NSF award PHY-1306702. We thank James Clark for his thoughtful comments.

\bibliography{chirp.bib}

\end{document}